\documentclass[prc,aps,twocolumn,showpacs]{revtex4}
\usepackage{graphicx}
\begin{document}

\title{Many-body fits of phase-equivalent effective interactions}

\author{Calvin W. Johnson}
\affiliation{Department of Physics, San Diego State University,
5500 Campanile Drive, San Diego, CA 92182-1233}


\begin{abstract}
In many-body theory it is often useful to renormalize short-distance, high-momentum components of an interaction via unitary transformations. 
Such transformations preserve the on-shell physical observables 
of the two-body system (mostly phase-shifts, hence unitarily-connected effective interactions are often called phase-equivalent), while modifying off-shell T-matrix elements
influential in  many-body systems.   
In this paper I lay out a general and systematic approach for controlling the off-shell behavior of an effective interaction, which can be adjusted to many-body properties, and present an application to trapped fermions at the unitary 
limit.  
\end{abstract}
\maketitle

The force between two particles has to be determined empirically, but some matrix elements
 are easier to determine than 
others. For an isolated two-body system one can measure elastic phase shifts and bound state 
eigenvalues, but not off-shell (inelastic) matrix elements (specifically, the 
elements of the T-matrix).  Such matrix elements show up
when the two-body 
system is not isolated, i.e., embedded in a many-body system, but disentangling 
the off-shell two-body matrix elements from many-body data is not possible. 

Following modern trends in effective interaction theory \cite{OLS,UCOM,SRG,ABF08,Stetcu07}, 
I will argue that one can flip 
the ambiguity of off-shell matrix elements into an advantage. This paper illustrates
a general method to find the unitary transformation that preserves 
two-body observables while simultaneously providing the best fit to many-body data.

For example, despite much effort we do not have a unique prescription for the force 
between nucleons. Nonetheless high-precision data on two-nucleon systems \cite{SKR93}
strongly constrain any description, leading to a variety of competing interactions 
\cite{nijmegen,AV18,CDBonn,EHM09} which all, by construction, have indistinguishable  on-shell 
T-matrix elements; these are called phase-shift equivalent potentials. The off-shell T-matrix elements, which \textit{do} differ, can only be compared through many-body observables, including binding energies and excitation spectra.



In a parallel vein, consider configuration-interaction (CI) many-body calculations, 
which have several advantages but are vulnerable to strong short-range,
high-momentum components of the interaction, because
practical considerations require truncation, 
indicated by a projection operator $\hat{P}$. The problem is 
 the eigenvalues of $\hat{P}\hat{H} \hat{P}$ are not the same as the eigenvalues 
of $\hat{H}$ in an infinite or even very large space, and converge slowly with increasing 
dimension of $\hat{P}$, a particularly severe problem for CI calculations. 
Therefore one turns to effective interactions, generated using a unitary 
transformation 
\begin{equation}
\hat{H}_\mathrm{eff} =\hat{U}^\dagger \hat{H} \hat{U}. \label{Heff}
\end{equation} 
The hope is 
that the eigenvalues of $\hat{P}\hat{H}_\mathrm{eff} \hat{P}$ in the truncated 
space converge to those of $\hat{H}$ in an infinite space. 
 One common strategy is the so-called 
`cluster approximation': forcing the eigenvalues in the truncated space to agree 
with the `exact' values of the original space, but only for two- (or on occasion three-) 
particle systems, and then apply to many-body systems. 

Modern, rigorous effective interaction methods explicitly or implictly apply a unitary
transform that leave the on-shell behavior unchanged but which dial away the troublesome high-momentum part of the interaction. These methods, which include but are not limited to Okubo-Lee-Suzuki (OLS) \cite{OLS}, the unitary correlation 
operator method (UCOM) \cite{UCOM}, the similarity renormalization group (SRG) \cite{SRG},   and Alhassid-Bertsch-Fang (ABF) \cite{ABF08}, 
 generate a portfolio of new `phase-equivalent' interactions 
with the same on-shell T-matrices but different off-shell matrix elements.



The two-body cluster approximation is  complicated by the fact 
that three-body forces naturally arise out of effective field theory \cite{EHM09,SW92}. 
In practice, however, 
three-body forces make for computational difficulties. Thus several authors have sought to 
\textit{minimize} three-body forces by exploiting the interplay between 
three-body forces and off-shell matrix elements \cite{UCOM,INOY,JISP16}. 

This paper combines several of these ideas. 
 Borrowing  from phenomenological fitting of interactions \cite{BG77,BR06}, 
I show how one can
 choose a general unitary transformation to generate a phase-equivalent potential which 
 is a ``best fit'' to many-body data.  

%
%

To illustrate, I take on a specific challenge:  spin-1/2 fermions 
at the unitary limit, the so-called ``Bertsch problem,'' in the context 
of an external harmonic trap \cite{BertschProblem,BSG07}. 
The interaction is zero range and has infinite scattering length, similar to
the short-range, large scattering length nuclear force. 
The Hamiltonian is 
\begin{equation}
\hat{H} = \sum_i -\frac{\hbar^2 }{2m}\nabla^2_i + \frac{1}{2}m\Omega^2 r^2_i 
- V_0 \sum_{i <j} \delta^{(0)} \left ( \vec{r}_i - \vec{r}_j \right).
\label{hamiltonian}
\end{equation}
 The ground state energy for the 3-body 
case is known analytically \cite{WC06}, 
and for $N \geq 4$ I take as `exact' the results from correlated 
Gaussian and fixed-node diffusion Monte Carlo calculations \cite{BSG07} (with
quoted statistical and systematic errors of only a few percent, which 
I leave out).  Table I lists
the energies adopted in this study.  For purposes of comparison I 
group the energies into two sets: Set I includes the ground states and selected 
excited states for $N=3,4$, while Set II is comprised of the ground state energies 
for $N=3$-$10$. 

\begin{table}[h!]
\caption{Adopted energy levels \cite{BSG07,WC06} in units of 
$\hbar \Omega$ of the trapping potential and assignment to 
comparison sets.}
\begin{tabular}{|c|cc|c|}
\colrule
$N$ & $L^\pi;S$ &  energy  & Set \\
\colrule 
3 & $1^-; \frac{1}{2}$ & 4.27 & I,II \\
  & $0^+; \frac{1}{2}$ &   4.66 & I \\
\colrule
4 & $0^+;0$  & 5.05 & I,II\\
& $2^+;0$ &  5.91  & I	\\
& $1^+;1$ & 6.58 & I \\
\colrule
5 & $1^-;\frac{1}{2}$ & 7.53 & II \\
\colrule
6 & $0^+;0$  & 8.48&  II \\
\colrule
7 & $1^+;\frac{1}{2}$ & 11.36 & II \\
\colrule
8 & $0^+;0$ & 12.58 & II \\
\colrule
9 & $0^+;\frac{1}{2}$ & 15.69 & II \\
\colrule
10 &$0^+;0$ & 16.80 &  II \\
\colrule
\end{tabular}
\end{table}

The  Hamiltonian (\ref{hamiltonian}) can be separated into  
the center of mass  plus the relative Hamiltonian:
\begin{equation}
\hat{H}_\mathrm{rel} =-\frac{\hbar^2}{2\mu} \nabla^2_{r}
+ \frac{1}{2}\mu\Omega^2 r^2 -V_0 \delta \left ( r \right)
\end{equation}
where $\vec{r} = \vec{r}_1 - \vec{r}_2$ is the relative coordinate and 
$\mu = m/2$  the reduced mass. 
For application to configuration interaction (CI) 
calculations, one computes
the matrix elements of $\hat{H}_\mathrm{rel}$ in a harmonic oscillator 
basis,  
$ \langle n^\prime l | \hat{H}_\mathrm{rel} | n l \rangle$ 
 (which are only nontrivial for the relative $s$-wave or $l=0$), 
and then transform to the lab frame two-body matrix elements 
via Brody-Moshinsky brackets \cite{BM60}.


I  truncated the
relative space to $n_\mathrm{cutoff}=5$, which means for 
$ \langle n^\prime l | \hat{H}_\mathrm{rel} | n l \rangle$  I use 
$n,n^\prime = 1,\ldots,5$; this corresponds 
to including up $8 \hbar \Omega$ in excitation energy in the relative space. 
(Calculations with different $n_\mathrm{cutoff}$ had  similar results.) 
For the lab frame, also using a harmonic oscillator basis, I truncated the single-particle space to four major shells,  that is,  $0s$, $0p$, $1s0d$, and $1p0f$.

The $\delta$-interaction must be regularized to fix the 
scattering length; for CI calculations in an oscillator basis, the 
interaction strength $V_0$ depends on 
$n_\mathrm{cutoff}$, the number of $s$-wave basis states used . 
For the `bare' interaction, fixing the 
scattering length is equivalent to fixing the ground state energy of the relative 
two-body state \cite{ABF08,Stetcu07,BERW98}; 
due to truncation, the excitation energies  differ 
from the infinite space values.
 Alternately, I also used, in the $n_\mathrm{cutoff} =5$ 
space, the ABF interaction, which fixes all five eigenenergies of $\hat{H}_\mathrm{rel}$ 
in the truncated space to the correct (infinite space) values. 

Going to three or more particles immediately illustrates  some of the headaches of effective interactions, as can be seen in Table II. 
Consider the rms error between calculated and target (experimental) energies,
\begin{equation}
[\sum_\alpha (E_\alpha(\vec{c}) - E^0_\alpha)^2]^{1/2} \label{rms}.
\end{equation}
For 'bare' interactions, the rms error on Set II (ground state 
energies, in units of $\hbar \Omega$ of the trapping potential,
for $A$=3-10) is 1.16, while for ABF it is 2.32. 
That is, forcing the effective 
interaction to have the correct eigenvalues for the two-body system can lead to larger 
errors in the many-body system. (This comparison, oversimplifies the story;  
please read the original \cite{ABF08}.)

\begin{table}[h!]
\caption{Root-mean-square error between adopted exact energies (Table I) 
and calculated CI energies. `Starting $\hat{H}_\mathrm{rel}$' refers 
to using either the 'bare' or ABF regularized interaction (see text).  
`Generators' refer to the set of operators 
used in the unitary transformation to minimize the rms error: 
`none' means no transformation was performed, `all' means all ten 
generators were used, and $d/dr$  denotes using 
only the single generator $d/dr$ in the relative space.  
Units are $\hbar \Omega$ of trapping potential. The unitary transformations 
were fit to either Set I or II; `$I\rightarrow II$' refers to fitting to Set I, but computing the rms error on Set II. 
}
\begin{tabular}{|c|c||c|c|c|}
\colrule
Starting  & Fit & \multicolumn{3}{|c|}{ Generators} \\
\cline{3-5}
$\hat{H}_\mathrm{rel}$ & Set  & none & $d/dr$ & all \\
\colrule
bare & I & 0.62 & 0.19 & 0.10 \\
\colrule
bare & I$\rightarrow$II & (1.16) & 0.55 & 0.32 \\
\colrule
bare & II & 1.16 & 0.31 & 0.28 \\
\colrule
ABF & I & 1.06 & 0.11 & 0.06 \\
\colrule
ABF & I$\rightarrow$II & (2.32) & 0.58 & 0.37 \\
\colrule
ABF & II & 2.32 & 0.26 & 0.25 \\
\colrule
\end{tabular}


\end{table}

The apparent paradox of a bare interaction doing better than a renormalized one can be  understood by writing $\hat{U} = \exp(\hat{A})$,
where $\hat{A}$ is an anti-Hermitian two-body operator. 
Then $\hat{U}$ induces many-body terms in $\hat{H}_\mathrm{eff}$ cutoff by the 
two-body cluster approximation.


Previous authors made specific implicit choices in the form of their off-shell behavior 
to minimize the effect of three-body interactions. 
The strength of the correlation operator in UCOM \cite{UCOM}, which 
shifts nucleons away from each other, is fit to minimize the error in the $A=3$ and 
$4$ ground state energies.
The inside-nonlocal, outside-Yukawa (INOY) potential \cite{INOY} is adjusted to both 
two-body data plus the triton binding energy, while the $J$-matrix inverse scattering 
potential (JISP16) is fit to two-body data plus binding energies up to $^{16}$O \cite{JISP16}. 

I propose a  general method to choose the 'best'  unitary transformation.
Expand $\hat{A}$  in a  set 
of generators
\begin{equation}
\hat{A} = \sum c_i \hat{A}_i. \label{expandA}
\end{equation}  
The dependence 
of  the calculated many-body eigenenergies 
 $\{ E_\alpha(\vec{c}) \}$ on the $c_i$ can be expanded to first order 
\begin{equation}
E_\alpha(\vec{c}) \approx E_\alpha(0) + \sum_i \frac{\partial E_\alpha}{\partial 
c_i} c_i
\end{equation}
The required derivatives can be found via the Hellman-Feynman theorem \cite{F39}:
\begin{equation}
\frac{\partial E_\alpha}{\partial c_i} = \left \langle \alpha \left | 
\left [ \hat{H}, \hat{A}_i \right ] \right | \alpha \right \rangle \label{HF}
\end{equation}

Now consider a target set of (experimental) many-body energies  
$\{ E^0_\alpha\}$. By minimizing the root-mean-square difference 
(\ref{rms}) between calculated and target (experimental) energies,
one  obtains
\begin{equation}
\sum_j \left ( \sum_\alpha \frac{\partial E_\alpha}{\partial c_i} 
\frac{\partial E_\alpha}{\partial c_j}\right ) c_j \approx \sum_\alpha (E_\alpha^0
-E_\alpha ) \frac{\partial E_\alpha}{\partial c_i} \label{linearminimize}
\end{equation}
which can be solved for the $c_i$. Because Eq.~(\ref{linearminimize}) 
is only a linear approximation, one may need to iterate to converge on a 
best solution. This methodology is very similar to that used to 
fit two-body matrix elements to 
low energy nuclear spectra \cite{BG77,BR06}, except that in those cases one directly 
fits Hamiltonian matrix 
elements in the lab frame, and here I fit the coefficients of generators of a unitary transformation (and in my application below I work in the relative frame). 

(As an important technical point, the matrix $M_{ij} =\sum_\alpha \frac{\partial E_\alpha}{\partial c_i} 
\frac{\partial E_\alpha}{\partial c_j}$ is in general nearly singular, 
so one uses singular value 
decomposition \cite{MC,NR} to find the dominant modes. Here singular value 
decomposition is nothing more than using a spectral (eigenvalue) representation of 
$M_{ij}$ and keeping only nonsingular terms. )

This methodology is flexible and general. One can use as many or as few 
generators as desired or thought physically relevant, 
and can constrain to a best fit using an arbitrary choice 
of many-body energies.

\begin{figure}  
\includegraphics [width = 7.5cm]{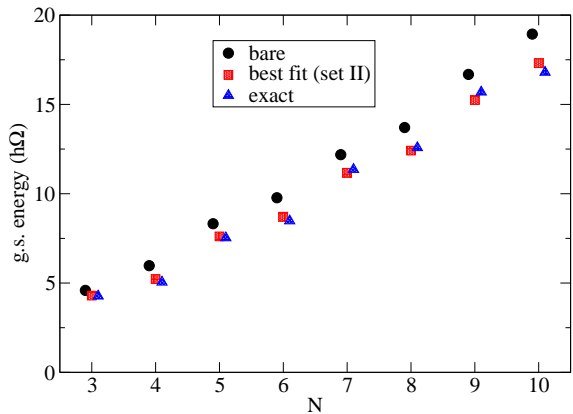}
\caption{\label{bestfit} (Color online) Comparison of ground state energy for $N$ particles 
in a harmonic trap, interacting via a contact interaction: (black) circles are 
``bare'' results (fixed in relative frame at $n_\mathrm{cutoff} = 5$), (red) squares 
are the best fit of all generators of a unitary transformation, 
and (blue) triangles are the exact results. 
All energies in units of $\hbar \Omega$ of the trapping potential. 
Slight horizontal offshifts were introduce to aid visibility.
}
\end{figure}

I applied this prescription to my example system. As described above, in the relative space
$ \langle n^\prime,l=0  | \hat{H}_\mathrm{rel} | n,l=0  \rangle$
 with $n_\mathrm{cutoff}=5$ is a $5 \times 5$ 
real symmetric matrix (using either the `bare' or ABF matrix elements).  I then introduced a general real orthogonal transformation as in Eq.~(\ref{Heff}), 
where $\hat{U}$ also is a $5\times 5$ matrix , with 10 possible antisymmetric generators
$\hat{A}_i$. (I only performed the transformation in the 
relative $s$ channel. In principle an effective interaction could generate nontrivial 
matrix elements in other channels. I leave this to future investigations.)
I then fit the parameters $c_i$ in Eq.~(\ref{expandA}) 
by minimizing the rms error on either Set I or Set II. The results
are shown in Table II (the energies are all in units of $\hbar \Omega$ of the trapping 
potential) and illustrated in Figure 1.  There is a dramatic reduction in the rms error.

The best fit parameters for Sets I and II differed, of course. I also considered the 
error in extrapolation, by fitting to Set I but calculating the rms error on
on the larger Set II; found in Table II in the rows marked 'I$\rightarrow$II,' 
the error in extrapolation is at worse fifty percent larger than the best fit to 
Set II.  (The first number in these rows, 
in parentheses, is the original rms error, as there is no fit.)

Interestingly, although ABF starts with a larger rms error, it yields fits with 
smaller errors.

By insisting that the on-shell two-body matrix elements (relative-space 
 eigenvalues) remain invariant, and keeping only the $s$-wave relative channel, 
the above is the best fit that one can obtain. The 
remaining residual error is due to induced three-body forces that cannot be fully 
replicated using a two-body force \cite{BJ09}.

The SVD of $M_{ij}$ was always dominated by one or two eigenvalues and thus one or 
two generators. In addition, one would appreciate simple physical insights into the 
unitary transformation. 
I therefore tried the manifestly antisymmetric generator   
$d/dr$ (where $r$ here is 
the relative coordinate),  because of simplicity but also because of 
similarity to UCOM \cite{UCOM}.  Table II shows  this single generator  
significantly reduces the rms error. 


The minimization will be model-space 
dependent, but one can see this as either a weakness or a strength. 
Most many modern interactions already have an intrinsic cutoff dependence.  
Here I have given how to best ``tune''
an effective interaction to a model space. One can also choose key energy levels 
to fit to, not only ground state energies but excited states that contain important 
physics such as spin-orbit splitting. Also, critically, it will be important to 
constrain the unitary transformation to other observables such as the rms radius. 
Given that only a few degrees of freedom were used to minimize the energies, 
this is not impossible and is under investigation.


To summarize, I have discussed general unitary transformations which produce effective 
interactions, and have shown how one can 
use many-body data to improve the interaction while 
preserving eigenvalues in the relative space (on-shell T-matrix elements).
This methodology is a generalization of modern effective interaction theory 
and of previous specific attempts to reduce the need for three-body interactions.
With a simply 
defined yet numerically challenging case of trapped fermions at the unitary limit, I demonstrated 
 improvement in ground state energies. 

One of the claims of current methods is  they provide reliable error estimates 
for predictions. While I have not addressed this important issue, one should see 
that the error estimates are dependent on the choice of unitary transformation, 
an issue that has not yet been addressed in a deep way.  At least 
I hope to provokes a closer investigation of competing 
effective interaction methodologies (and their claimed error estimates): OLS, SRG, UCOM, and other  specific choices. 

Extending this work to nuclei and to the inclusion of observables is under way.

The U.S.~Department of Energy supported this investigation through
grants DE-FG02-96ER40985 and, under the auspices of the Universal Nuclear Energy 
Density Functional project, DE-FC02-09ER41587.  I thank M.~Bromley and I.~Stetcu for helpful feedback 
on the manuscript.

\end{document}